\documentclass[a4paper,10pt]{article}

\usepackage[utf8]{inputenc}
\usepackage[T1]{fontenc}

\usepackage{a4wide}

\usepackage{amsmath,amssymb}
\usepackage{amsthm}
\usepackage{mathtools}
\mathtoolsset{showonlyrefs=true} 
\usepackage{stackengine} 
\stackMath

\usepackage{paralist} 

\usepackage{authblk}

\usepackage[dvipsnames,table,xcdraw,svgnames]{xcolor}
\definecolor{highlightNEW}{named}{black}
\definecolor{mylilas}{RGB}{170,55,241}
\definecolor{mygreen}{RGB}{70,180,5}
\definecolor{myred}{RGB}{244,63,43}

\usepackage{xurl} 
\usepackage{hyperref}
\hypersetup{
    breaklinks=true,
    colorlinks=true,
    linkcolor=Navy,
    citecolor=Navy,
    urlcolor=Navy
}

\usepackage{graphicx}
\graphicspath{{figs/}}
\ifpdf
\usepackage{epstopdf}
\fi
\usepackage{subcaption}
\usepackage[section]{placeins} 

\usepackage{booktabs}
\usepackage{multirow}

\usepackage[absolute,overlay,showboxes]{textpos}
\TPMargin{2mm}
\textblockrulecolour{Navy}

\usepackage{natbib}

\usepackage{listings}
\usepackage{soulutf8}
\sethlcolor{LimeGreen}

\soulregister\cite7%
\soulregister\citep7%
\soulregister\eqref7%
\soulregister\pageref7%
\soulregister\ref7%

\newtheorem{theorem}{Theorem}[section]

\newtheorem{remark}[theorem]{Remark} 

\newcommand{\doi}[1]{DOI~\href{\detokenize{http://dx.doi.org/#1}}{\detokenize{#1}}}
\newcommand{\zblnumber}[1]{Zbl~\href{\detokenize{https://zbmath.org/?q=an:#1}}{\detokenize{#1}}}
\newcommand{\mrnumber}[1]{\href{\detokenize{https://www.ams.org/mathscinet-getitem?mr=#1}}{\detokenize{MR#1}}}

\renewcommand{\d}{\,\mathrm{d}}
\newcommand{\dx}{\,\mathrm{d}x}

\newcommand{\e}{\mathrm{e}}
\newcommand{\E}{\mathbb{E}}

\newcommand{\R}{\mathbb{R}}

\newcommand\ntilde[2][2]{%
 \def\useanchorwidth{T}%
  \ifnum#1>1%
    \stackon[-1.3ex]{\ntilde[\numexpr#1-1\relax]{#2}}{\mathchar"307E\kern-.5pt}%
  \else%
    \stackon[-1ex]{#2}{\mathchar"307E\kern-.5pt}%
  \fi%
}

\newdimen\CdotAxis
\newcommand*{\CdotAux}[3]{%
  {%
    \settoheight\CdotAxis{$#2\vcenter{}$}%
    \sbox0{%
      \raisebox\CdotAxis{%
        \scalebox{#1}{%
          \raisebox{-\CdotAxis}{%
            $\mathsurround=0pt #2#3$%
          }%
        }%
      }%
    }%
    \dp0=0pt %
    \sbox2{$#2\bullet$}%
    \ifdim\ht2<\ht0 %
      \ht0=\ht2 %
    \fi
    \sbox2{$\mathsurround=0pt #2#3$}%
    \hbox to \wd2{\hss\usebox{0}\hss}%
  }%
}
\makeatletter
\def\mathcolor#1#{\@mathcolor{#1}}
\def\@mathcolor#1#2#3{%
  \protect\leavevmode
  \begingroup
    \color#1{#2}#3%
  \endgroup
}
\makeatother

\newcommand{\ccode}[2]{\par
        \vspace*{8pt}
        {{\leftskip18pt\rightskip\leftskip
        \noindent{\it #1}\/: #2\par}}\par}

\makeatletter
\@ifundefined{keywords}{\newcommand{\keywords}[1]{\ccode{Keywords}{#1}}}{}
\@ifundefined{email}{\newcommand{\email}[1]{\href{mailto:#1}{#1}}}{}
\makeatother

\usepackage[mathscr]{euscript}
\DeclareSymbolFont{rsfs}{U}{rsfs}{m}{n}
\DeclareSymbolFontAlphabet{\mathscrsfs}{rsfs}

\def\received#1{Received~#1\par}
\def\revised#1{Revised~#1\par}

\newcommand{\jpTitle}{A note on a PDE approach to option pricing under xVA}

\newcommand{\jpKeywords}{value adjustment; PDE; collateral; Monte-Carlo simulation}
\newcommand{\jpMSC}{91G20; 91B70; 91G40}%
\newcommand{\jpJEL}{G12; C63}%
\newcommand{\jpDateReceived}{30 April 2021} 
\newcommand{\jpDateRevised}{20 July 2021}
\newcommand{\jpDate}{}

\author[1]{Falko Baustian} 
\author[2]{Martin Fencl}
\author[2]{Jan Posp\'{\i}\v{s}il\thanks{Corresponding author, \email{honik@kma.zcu.cz}}} 
\author[2]{Vladim\'{\i}r \v{S}v\'{\i}gler}
\affil[1]{Department of Mathematics, University of Rostock, Ulmenstra\ss e 69, 18057 Rostock, Germany}
\affil[2]{NTIS - New Technologies for the Information Society, Faculty of Applied Sciences, \authorcr University of West Bohemia, Univerzitn\'{\i} 2732/8, 301 00 Plze\v{n}, Czech Republic,\vspace*{3pt}}

\title{\textcolor{Navy}{\textsc{\jpTitle}}}
\date{\jpDate}

\begin{document}

\maketitle

\begin{center}
\received{\jpDateReceived}
\revised{\jpDateRevised}
\end{center}

\begin{abstract}
In this paper we study partial differential equations (PDEs) that can be used to model value adjustments. Different value adjustments denoted generally as xVA are nowadays added to the risk-free financial derivative values and the PDE approach allows their easy incorporation. The aim of this paper is to show how to solve the PDE analytically in the Black-Scholes setting to get new semi-closed formulas that we compare to the widely used Monte-Carlo simulations and to the numerical solutions of the PDE. Particular example of collateral taken as the values from the past will be of interest.

\end{abstract}

\ccode{MSC classification}{\jpMSC}
\ccode{JEL classification}{\jpJEL}

\setcounter{tocdepth}{2}
\tableofcontents
\clearpage

\section{Introduction}\label{sec:introduction}

The famous Black-Scholes model for option pricing in an arbitrage-free environment was introduced in 1973 by \cite{BlackScholes73} and \cite{Merton73}. The pioneering model was very popular in the beginning and widely used by practitioners. One of the basic assumptions of the Black-Scholes model is that the volatility of the underlying asset is constant. This hypothesis was later discarded after \cite{Rubinstein85} observed that the real-market prices of out-of-the-money call options with short maturity were significantly higher than the predictions of the model. \cite{HullWhite87} introduced a model with stochastic volatility to fix this issue. The stock market crash of October 1987 with the infamous Black Monday permanently changed the shape of the volatility surface, see \cite[Chapter 18.3]{Hull09}. After the crash the implied volatility developed the marked volatility smile as observed today that \cite{Rubinstein94} explained with the fear of traders regarding the possibility of a similar event. As a consequence, stochastic volatility models by \cite{HullWhite87}, \cite{SteinStein91}, \cite{Heston93}, and others became the standard in option pricing. The financial crisis that started in 2007 substantially changed the rules for derivative pricing again. The possibility to borrow and lend any amount of money at a risk-free interest rate is a crucial assumption of all the above mentioned models since its standardly used in the hedging arguments for the price valuation of the options. As the crisis culminated in the bankruptcy of Lehman Brothers and other big financial entities this conjecture became obsolete. Since the default of a trading partner is not in the scope of the classic option pricing models the traders had to react to this shortage in the models and included a collateral against default in many trades. Both parties of a contract have to consider the potential default of the counterparty. The traders respond to this bilateral counterparty risk with credit value adjustments (CVA), \cite{SorensonBollier94} and \cite{JarrowTurnbull95}, to have a collateral against default of the counterparty. The corresponding collateral of the counterparty, e.g., a bank, is covered by a debit value adjustment (DVA), \cite{DuffieHuang96}. In addition to this adjustments, derivative contracts can also contain a collateral, also called a collateral value adjustment (COLVA), regulated by the credit support annex (CSA). For more details on risk management see \cite{Hull18Risk}. Convexity adjustments that describe the differences between the prices of collateralized and non-collateralized derivatives were studied by \cite{Piterbarg10}. Funding cost adjustments (FCA) are other common price adjustments that traders use to include funding cost in the price of the contract. According to \cite{BurgardKjaer11} and \cite{HullWhite12} FCA have no theoretical basis. Since traders use them anyway \cite{HullWhite14} studied the adjustments with focus on their practical realization. There exist also other types of adjustments like capital value adjustments (KVA), margin value adjustments (MVA), and liquidity value adjustments (LVA) but we will restrict ourselves just to CVA, DVA, and FCA. The various adjustments to the value of the derivative are subsumed as xVA. A comprehensive guide to xVA is the book by \cite{Gregory2020}.

In analogy to the Black-Scholes approach, \cite{BurgardKjaer11PDE} derived a partial differential equation (PDE) for the adjusted derivative price under bilateral counterparty risk and funding costs. The partial differential equation is linear if the mark-to-market value at default is considered to be the non-default value of the derivative and nonlinear if the mark-to-market value is given by the adjusted value of the derivative with all value adjustments. The latter case is less meaningful from the practitioner's point of view since the close out is based in the risk-free mark-to-market for contracts that follow the 2002 International Swaps and Derivatives Association (ISDA) Master Agreement\footnote{available online at \url{https://www.isda.org/book/2002-isda-master-agreement-english/} [cit. 2019-09-12].}. Hence, we will study the linear partial differential equation introduced in \cite{BurgardKjaer13} which takes collaterals into account.

Value adjustments are normally applied to portfolios with various derivatives. Nevertheless, we consider a single option contract to analyze the influence of the different adjustments. The macro view, i.e., the homogeneous cost view for the whole portfolio, essentially follows from the micro view of single contracts. We derive the analytic solution of the PDE of \cite{BurgardKjaer13}, obtain a semi-closed pricing formula for the value adjustments, and compare the formula to widely used Monte-Carlo (MC) simulations and numerical solutions of the partial differential equations as in \cite{Arregui17} and \cite{Arregui18}. The problem has recently been studied for the Heston model in \cite{SalvadorOosterlee21}.

The paper is structured in the following way. We introduce the PDE for the correction between the adjusted price with xVA and the risk-free price of a derivative contract and we discuss various collaterals in Section~\ref{sec:preliminaries}. In Section~\ref{sec:methodology}, we introduce the pricing formula and give details on the numerical schemes, before numerical results are presented in Section~\ref{sec:results}. We conclude in Section~\ref{sec:conclusion}.

\section{Preliminaries}\label{sec:preliminaries}

We consider a generic derivative contract (e.g. option), possibly collateralized, between an issuer $\text{B}$ and a counterparty $\text{C}$, with an economic value $\check{V}$ that incorporates the risk of default of counterparty and issuer and any net funding costs the issuer may encounter prior to own default. We are mainly interested in the correction $U=\check{V}-V$ to the risk-free Black-Scholes price $V$. Using the Black-Scholes PDE for $V$ one can derive \citep{BurgardKjaer13} the corresponding PDE for $U$.

Let $\mathcal{A}$ be the standard Black-Scholes operator
\begin{equation}
\mathcal{A} = \frac{1}{2}\sigma^2 S^2 \frac{\partial^2}{\partial S^2}+(q_S-\gamma_S)S \frac{\partial}{\partial S},
\end{equation}
where $\sigma>0$ is the constant volatility and $q_S-\gamma_S$ is the effective financing rate of the market factor hedge $S$. We will study the following final value problem for function $U(t,S)$

\begin{equation}
\left\{
\begin{aligned}
\frac{\partial U}{\partial t}(t,S) + \mathcal{A}U(t,S) - (r+\lambda_\text{B}+\lambda_\text{C})U(t,S) 
&= f(V(t,S)), &&\qquad\text{ for } t\in[0,T), S\geq 0, \label{e:PDE_U} \\
U(T,S) &= 0, &&\qquad\text{ for } S\geq 0.
\end{aligned}
\right.
\end{equation}
where the inhomogeneity $f$ has the form
\begin{align*}
f &= f_\text{CVA} + f_\text{DVA} + f_\text{FCA} + f_\text{COLVA}, \\
f_\text{CVA} &= -\lambda_\text{C}(g_\text{C}-V), \\
f_\text{DVA} &= -\lambda_\text{B}(g_\text{B}-V), \\
f_\text{FCA} &= \lambda_\text{B} \cdot \varepsilon_h, \\ 
f_\text{COLVA} &= s_X \cdot X,
\end{align*}
with constants $\lambda_\text{B},\lambda_\text{C},s_X$ and the hedge error $\varepsilon_h$ described below. Here, $r$ is the risk-free rate, $\lambda_\text{B},\lambda_\text{C}$ are the default intensities of the counterparties, $\varepsilon_h$ describes the size of the hedge error, and $s_X$ is the difference between the collateral rate and the risk-free rate. $X$ denotes the collateral and we have
\begin{align*}
g_\text{C} &= R_\text{C}(V-X)^{+}-(V-X)^{-}+X,\\
g_\text{B} &= (V-X)^{+}-R_\text{B}(V-X)^{-}+X, 
\end{align*}
where $R_\text{B},R_\text{C}\in[0,1]$ are the recovery rates of the derivative position. The hedge error is given by $\varepsilon_h=g_\text{B}+P_D-X$, where $P_D$ is the value of the issuers bond portfolio in case of default of the issuer. Hence, different semi-replication strategies result in different hedging errors, e.g., the hedge error vanishes for perfect replication. We consider a regular bilateral close-out for a semi-replication strategy that has no shortfall at default of the issuer. In this case the hedge error $\varepsilon_h$ becomes
\begin{equation}\label{e:varepsilon}
\varepsilon_h=(1-R_B)(V-X)^+.
\end{equation}
A discussion of different semi-replication strategies can be found in \cite{BurgardKjaer13}. The positive and negative part of a function $\varphi$ are defined as $\varphi^+=\max(\varphi,0)$ and $\varphi^-=\max(-\varphi,0)$, respectively. We remark that \cite{BurgardKjaer13} use the definition $\min(\varphi,0)$ for the negative part resulting in slightly different equations.\\\\
We are interested in different standard types of collaterals. For un-collateralized derivative contracts, we have $X=0$ and the terms $f_\text{CVA}$, $f_\text{DVA}$ and $f_\text{FCA}$ simplify to
\begin{equation*}
f_\text{CVA}=\lambda_\text{C}(1-R_\text{C})V^+, \quad f_\text{DVA}=-\lambda_\text{B}(1-R_\text{B})V^-, \quad f_\text{FCA}=\lambda_\text{B}(1-R_\text{B})V^+.
\end{equation*}
In one-way CSA, the issuer posts collateral provided that the risk-free value is out-of-the-money. That corresponds to $X=-V^-$ and we get
\begin{equation*}
f_\text{CVA}=\lambda_\text{C}(1-R_\text{C})V^+, \quad f_\text{DVA}=0, \quad f_\text{FCA}=\lambda_\text{B}(1-R_\text{B})V^+.
\end{equation*}
For gold-plated two-way CSA with collateral $X=V$ there are no CVA, DVA, and FCA. If the collateral rate equals the risk-free rate then the term $f_\text{COLVA}$ vanishes and the adjusted price $\check{V}$ of a two-way CSA becomes the risk-less price $V$.

In real world situations, there is most likely a collateral that depends on $V$ non-smoothly, non-linearly or even discontinuously. For further reading about these dependencies we refer the reader for example to ISDA: Legal Guidelines for Smart Derivatives Contracts: Collateral\,\footnote{available online at \url{https://www.isda.org/2019/09/12/legal-guidelines-for-smart-derivatives-contracts-collateral/} [cit. 2019-09-12].}.

A popular choice for a collateral among practitioners is the \emph{value from the past}, i.e. $X = V(t-t_0,S(t-t_0))$, where $S(t-t_0)$ is the value of the underlying asset price from the past and $t_0$ is the \emph{margin period of risk}, commonly assumed to be 10 business days for daily margining. With the notation $\Delta_t V=V(t,S(t))-V(t-t_0,S(t-t_0))$ we obtain
\begin{align*}
f_\text{CVA} &= \lambda_\text{C}(1-R_\text{C})(\Delta_t V)^+, \\
f_\text{DVA} &= -\lambda_\text{B}(1-R_\text{B})(\Delta_t V)^-, \\
f_\text{FCA} &= \lambda_\text{B}(1-R_\text{B})(\Delta_t V)^+, \\
f_\text{COLVA} &= s_X V(t-t_0,S(t-t_0)),
\end{align*}
for this collateral. Typically $t$ denotes the current time, $t_0=10/252$ represents ten business days. If $V$ is the Black-Scholes call option price, we have that
\begin{equation}\label{e:NPV} 
V(t-t_0,S(t-t_0)) = N(d_1(t-t_0,S(t-t_0))) S(t-t_0) + K\e^{-r(T-(t-t_0))} N(d_2(t-t_0,S(t-t_0)), 
\end{equation}
where $N(\cdot)$ is the standard normal cumulative distribution function and 
\[ d_{1,2}(s,x) = \frac1{\sigma\sqrt{T-s}} \left[ \ln\frac{x}{K} + \left( r \pm \frac12 \sigma^2\right)(T-s) \right]. \]

Applying the Feynman-Kac Theorem to PDE \eqref{e:PDE_U} gives us \citep{BurgardKjaer13} the semi-closed solution in the form
\begin{equation}
U = U_\text{CVA} + U_\text{DVA} + U_\text{FCA} + U_\text{COLVA}, \label{e:sol_MC}
\end{equation}
with
\begin{align}
U_\text{CVA} &= -\int_t^T \lambda_\text{C}(u)D_{r+\lambda_\text{B}+\lambda_\text{C}}(t,u)\E_t[V(u)-g_\text{C}(V(u),X(u))]\d u, \\
U_\text{DVA} &= -\int_t^T \lambda_\text{B}(u)D_{r+\lambda_\text{B}+\lambda_\text{C}}(t,u)\E_t[V(u)-g_\text{B}(V(u),X(u))] \d u,\\
U_\text{FCA} &= -\int_t^T \lambda_\text{B}(u)D_{r+\lambda_\text{B}+\lambda_\text{C}}(t,u)\E_t[\varepsilon_h(u)]\d u, \\
U_\text{COLVA} &= -\int_t^T s_X D_{r+\lambda_\text{B}+\lambda_\text{C}}(t,u)\E_t[X(u)]\d u,
\end{align}
where $D_{\hat{r}}(t,u) = \exp\left\{ -\int_t^u \hat{r}(v)\d v\right\}$ is the discount factor between $t$ and $u$ for a rate $\hat{r}$. The measure of the expectations in these equations is such that $S$ drifts at rate $(q_S-\gamma_S)$. The sum of DVA and FCA is sometimes referred to as funding valuation adjustments (FVA).

\section{Methodology}\label{sec:methodology}

In this section we present our main results, in particular we show how to analytically solve the corresponding PDE by a transformation to the heat equation and present a direct relation between the correction and the risk-less price for certain contracts with non-negative payoff. We discuss the boundary conditions for the numerical schemes and provide details on the numerics of all three used approaches: finite differences, the semi-closed solution formula, and the Monte-Carlo method.

We perform the standardly used change of variables to the time till maturity and the logarithm of the stock price
\begin{equation}\label{e:chov}
\tau=T-t\quad\text{ and }\quad x=\ln S,
\end{equation}
and introduce new functions $u(\tau,x)\cong U(t,S)$ and $v(\tau,x)\cong V(t,S)$. After applying \eqref{e:chov} to the PDE \eqref{e:PDE_U} we get
\begin{equation}
\frac{\partial u}{\partial \tau}=\frac{1}{2}\sigma^2 \frac{\partial^2 u}{\partial x^2}+\left(q_S-\gamma_S-\frac{1}{2}\sigma^2\right) \frac{\partial u}{\partial x}-(r+\lambda_\text{B}+\lambda_\text{C})u-f.
\end{equation}
If we set
\begin{equation}\label{e:rho}
\rho=q_S-\gamma_S-\frac{1}{2} \sigma^2
\end{equation}
to simplify the notation then we can write the initial value problem for the correction $u(\tau,x)$ as
\begin{equation}\label{e:PDE_u}
\left\{
\begin{aligned}
\frac{\partial u}{\partial \tau} &= \frac{1}{2}\sigma^2 \frac{\partial^2 u}{\partial x^2}+\rho\frac{\partial u}{\partial x}-(r+\lambda_\text{B}+\lambda_\text{C})u-f, &&\qquad\text{ for } \tau\in(0,T], x\in\R, \\
u(0,x) &= 0, &&\qquad\text{ for } x\in\R. 
\end{aligned}
\right.
\end{equation}
The PDE \eqref{e:PDE_u} is parabolic and has a unique weak solution \citep{Evans10}.

\subsection{Solution by transformation to heat equation}

Black and Scholes obtained their famous formula by transforming the PDE to the homogeneous heat equation from physics. If we want to get a heat equation from \eqref{e:PDE_u}, we have to perform successive transformations.

We start by introducing $\tilde{u}(\tau,x)=\e^{(r+\lambda_\text{B}+\lambda_\text{C})\tau}u(\tau,x)$ to simplify the PDE \eqref{e:PDE_u} to
\begin{equation}
\frac{\partial \tilde{u}}{\partial \tau} = \frac{1}{2}\sigma^2 \frac{\partial^2 \tilde{u}}{\partial x^2}+\rho\frac{\partial \tilde{u}}{\partial x}-f \cdot \e^{(r+\lambda_\text{B}+\lambda_\text{C})\tau}.
\end{equation}
Next, we use a suitable shift $\displaystyle\ntilde{u}(\tau,x)=\tilde{u}(\tau,x-\rho \tau)$ to obtain the \textit{non-homogeneous heat equation}
\begin{equation}
\displaystyle\frac{\partial \ntilde{u}}{\partial \tau} = \frac{1}{2}\sigma^2 \frac{\partial^2 \ntilde{u}}{\partial x^2}-f \cdot \e^{(r+\lambda_\text{B}+\lambda_\text{C})\tau}. \label{e:PDE_u_heat}
\end{equation}
We observe that $\displaystyle\ntilde{u}(0,x)=u(0,x)=0$ for all $x\in\R$ and we can write the solution of \eqref{e:PDE_u_heat} as
\begin{equation}
\displaystyle\ntilde{u}(\tau,x)=-\int_{0}^{\tau}\left(\int_{-\infty}^{\infty} \frac{1}{\sqrt{4\pi\frac{1}{2}\sigma^2(\tau-s)}} \e^{-\frac{(z-x)^2}{2\sigma^2(\tau-s)}} \cdot f(v(s,z))\d z\right) \e^{(r+\lambda_\text{B}+\lambda_\text{C})s}\d s.
\end{equation}
We perform one more substitution
$$
y = \frac{z-x}{\sqrt{2\sigma^2(\tau-s)}}
$$
to have the solution in the form
\begin{equation}
\displaystyle\ntilde{u}(\tau,x) = -\frac{1}{\sqrt{\pi}}\int_{0}^{\tau}\left(\int_{-\infty}^{\infty} \e^{-y^2} \cdot f(v(s,x+\sqrt{2\sigma^2(\tau-s)}y))\d y\right) \e^{(r+\lambda_\text{B}+\lambda_\text{C})s}\d s.
\end{equation}
Finally, we apply the inverse substitutions to obtain the formula for the solution of \eqref{e:PDE_u} in terms of the original variables
\begin{equation}\label{e:sol_heat}
u(\tau,x) = -\frac{1}{\sqrt{\pi}} \int_{0}^{\tau}\left(\int_{-\infty}^{\infty} \e^{-y^2} \cdot f(v(s,x+\rho\tau+\sqrt{2(\tau-s)}\sigma y))\d y\right)\e^{-(r+\lambda_\text{B}+\lambda_\text{C})(\tau-s)}\d s.
\end{equation}

\begin{remark}
Another representation of the same solution can be obtained by means of the Fourier transform
\begin{align}
\hat{u}(\tau,k):=\mathcal{F}\left[u(\tau,x)\right] &=\int_{-\infty}^{\infty}u(\tau,x) \e^{-i k x}\dx, \\
\hat{f}(\tau,k):=\mathcal{F}\left[f(v(\tau,x))\right] &= \int_{-\infty}^{\infty}f(v(\tau,x)) \e^{-i k x}\dx,
\end{align}
where $i$ denotes the complex unit. If we apply the Fourier transform to the PDE \eqref{e:PDE_u}, we get
\begin{equation}
\frac{\partial \hat{u}}{\partial \tau}
=-\frac{1}{2}\sigma^2 k^2 \hat{u}+i k\rho\hat{u}-(r+\lambda_\text{B}+\lambda_\text{C})\hat{u}-\hat{f}
=g(k)\hat{u}-\hat{f}, \label{e:PDE_hatu}
\end{equation}
where we denoted 
\begin{equation}\label{e:g}
g(k) = -\frac{1}{2}\sigma^2 k^2+ik\rho-(r+\lambda_\text{B}+\lambda_\text{C}). 
\end{equation}
We introduce $\hat{h}(\tau,k)=\e^{-g(k)\tau}\hat{u}(\tau,k)$ to transform \eqref{e:PDE_hatu} to the PDE
\begin{equation}
\frac{\partial \hat{h}}{\partial \tau} = -\e^{-g(k)\tau}\hat{f}
\end{equation}
that can be solved directly as an ordinary differential equation (ODE) in $\tau$ variable
\begin{equation}
\hat{h}(\tau,k) = -\int_{0}^{\tau} \e^{-g(k)s}\hat{f}(s,k)\d s.
\end{equation} 
Hence, the solution of \eqref{e:PDE_hatu} has the form
\begin{equation}
\hat{u}(\tau,k) = -\int_{0}^{\tau} \e^{g(k)(\tau-s)}\hat{f}(s,k)\d s 
\end{equation}
and we can apply the inverse Fourier transform to obtain the solution of the original equation \eqref{e:PDE_u}
\begin{align}
u(\tau,x) &= \mathcal{F}^{-1}\left[\hat{u}(\tau,k)\right] = \frac{1}{2\pi}\int_{-\infty}^{\infty}\hat{u}(\tau,k) e^{i k x}\d k \\
&= -\frac{1}{2\pi} \int_{-\infty}^{\infty} \e^{ikx} \left(\int_{0}^{\tau} \e^{(-\frac{1}{2}\sigma^2 k^2+ik\rho -(r+\lambda_\text{B}+\lambda_\text{C}))(\tau-s)}\hat{f}(s,k) \d s\right)\d k. \label{e:sol_fourier}
\end{align}
To show that two different forms \eqref{e:sol_heat} and \eqref{e:sol_fourier} of the solution are equivalent one may use for example similar techniques as \cite{BaustianMrazekPospisilSobotka17asmb}.
\end{remark}

\subsection{Non-negative payoffs}

Many derivative contracts like call options have a non-negative payoff resulting in a non-negative risk-less price $v$ of the derivative, i.e., we have $v=v^+$ and $v^-=0$. If we consider un-collateralized contracts or one-way CSA for such derivatives, as described in Section~\ref{sec:preliminaries}, then the term $f$ for the value adjustments simplifies to $f(v)=(\lambda_\text{C}(1-R_\text{C})+\lambda_\text{B}(1-R_\text{B}))v$ and we can represent the correction $u$ in dependence of the risk-less price $v$. We get
\begin{equation}\label{e:u_nn}
\left\{
\begin{aligned}
\frac{\partial u}{\partial \tau} &= \frac{1}{2}\sigma^2 \frac{\partial^2 u}{\partial x^2}+\rho\frac{\partial u}{\partial x}-(r+\lambda_\text{B}+\lambda_\text{C})u-c_fv, &&\qquad\text{ for } \tau\in(0,T], x\in\R, \\
u(0,x) &= 0, &&\qquad\text{ for } x\in\R. 
\end{aligned}
\right.
\end{equation}
with $c_f=(\lambda_\text{C}(1-R_\text{C})+\lambda_\text{B}(1-R_\text{B}))$ and recall that $v$ satisfies the Black-Scholes equation
\begin{equation}
\left\{
\begin{aligned}
\frac{\partial v}{\partial \tau} &= \frac{1}{2}\sigma^2 \frac{\partial^2 v}{\partial x^2}+\rho\frac{\partial v}{\partial x}-rv, &&\qquad\text{ for } \tau\in(0,T], x\in\R, \\
v(0,x) &= h(x), &&\qquad\text{ for } x\in\R, 
\end{aligned}
\right.
\end{equation}
where $h(x)\geq0$ is the payoff function of the derivative. We observe that the function $\frac{-c_fv}{\lambda_\text{B}+\lambda_\text{C}}$ is a solution of \eqref{e:u_nn} but has the non-trivial initial value
$$
\left(\frac{-c_fv}{\lambda_\text{B}+\lambda_\text{C}}\right)(0,x)=\frac{-c_f}{\lambda_\text{B}+\lambda_\text{C}}h(x).
$$
Fortunately, we can make use of the function $w=\e^{-(\lambda_\text{B}+\lambda_\text{C})\tau}v$ that satisfies $w(0,x)=h(x)$ and
\begin{equation*}
\frac{\partial w}{\partial \tau} = \frac{1}{2}\sigma^2 \frac{\partial^2 w}{\partial x^2}+\rho\frac{\partial w}{\partial x}-(r+\lambda_\text{B}+\lambda_\text{C})w
\end{equation*}
to obtain the solution
\begin{align*}
u(\tau,x) &= \frac{c_f}{\lambda_\text{B}+\lambda_\text{C}}(w(\tau,x)-v(\tau,x))
= \frac{\lambda_\text{C}(1-R_\text{C})+\lambda_\text{B}(1-R_\text{B})}{\lambda_\text{B}+\lambda_\text{C}}\left(\e^{-(\lambda_\text{B}+\lambda_\text{C})\tau}-1\right)v(\tau,x)
\end{align*}
of the initial value problem \eqref{e:u_nn} with trivial initial data. We can use this representation of the solution with any solution formula for $v$ like the Black-Scholes formula or a Fourier transform based formula to directly derive the correction term $u$.

\subsection{Boundary conditions}
\label{sec:meth_BC}
Both problems \eqref{e:PDE_U} and \eqref{e:PDE_u} are defined either on a half-line in $S$ or a line in $x$, respectively. However, if we want to compare the numerical solution of the problem \eqref{e:PDE_u} with the solution provided by MC or formula \eqref{e:sol_heat}, we must reduce ourselves to a finite interval in $x$. In such case, we need to define boundary conditions. 
In this section and the further text, we restrict ourselves to the European call options in order to prevent us from presenting duplicate ideas in the text.

The PDE in \eqref{e:PDE_U} is a good starting point here, rather than the PDE in \eqref{e:PDE_u}. Let us assume that we work on the finite interval $[0,S_{\max}]$ with $S_{\max}\in\mathbb{R}^{+}$ sufficiently large. For collaterals of the form $X=\alpha V^+ + \beta V^-$ with $\alpha,\beta\in\R$, we have $X(t,0) = V(t,0) = 0$ for all $t>0$ in which $V$ is the BS price of the European call option. Thus, if we set $S=0$, then the PDE in \eqref{e:PDE_U} reduces to
\begin{equation}
\frac{\partial U}{\partial t} = \left(r+\lambda_{\text{B}}+\lambda_{\text{C}}\right)\cdot U + \lambda_{\text{B}}\cdot\varepsilon_{h},
\end{equation}
which can be solved to get
\begin{equation}\label{e:S0_BC}
U(t,0) = \frac{\lambda_{\text{B}}\varepsilon_{h}}{r+\lambda_{\text{B}}+\lambda_{\text{C}}} \left( e^{\left(r+\lambda_{\text{B}}+\lambda_{\text{C}}\right) (t-T)} -1 \right).
\end{equation}
Hence, we have this Dirichlet boundary condition on the left for the PDE in \eqref{e:PDE_U}.\\
\indent Now, if we divide the PDE in \eqref{e:PDE_U} by $S^{2}$ and approach the limit $S\rightarrow +\infty$, we get
\begin{equation}
\lim\limits_{S\rightarrow +\infty} \frac{\partial^{2} U}{\partial S^{2}}(t,S) = 0.
\end{equation}

We are actually assuming here some behaviour of $U(t,S)$ as $S\rightarrow +\infty$. More precisely, the growth of $U(t,S)$ in $S$ should be sub-quadratic for large $S$. In pricing equations for $V$, a common approach is to consider the so-called \emph{zero gamma condition} \cite[Sec. 2.3]{Hout17}, i.e.,
\[ \frac{\partial^{2} V}{\partial S^{2}}(t,S_{\max}) = 0,\quad 0\leq t\leq T, \]
and hence it is natural to assume the same type of condition for the correction $U$
\begin{equation}\label{e:Sinf_BC}
\frac{\partial^{2} U}{\partial S^{2}}(t,S_{\max}) = 0, \quad 0\leq t\leq T,
\end{equation}
for some sufficiently large $S_{\max}$. This approach was also already used, e.g., by \cite{Castillo13} and \cite{Arregui17}. It is possible to assume here $U(t,S) = c_{0}(t) + c_{1}(t) S$ as it was done in these papers, substitute it into \eqref{e:PDE_U} and compute $c_{0},c_{1}$ either numerically or in some cases also analytically. However, we will stick with \eqref{e:Sinf_BC} as our boundary condition. It is easier to implement, does not require any additional lengthy computations and most importantly it provides us with significantly better results, when comparing the numerical solution of the PDE to Monte-Carlo (MC) or formula \eqref{e:sol_heat}.

Let us consider two numbers $x_{\min}, x_{\max}\in\mathbb{R}$ such that
\begin{equation}
-\infty< x_{\min} < 0 < x_{\max} < +\infty.
\end{equation}
Due to the transformation \eqref{e:chov} and derived boundary conditions \eqref{e:S0_BC},\eqref{e:Sinf_BC}, it seems reasonable to consider boundary conditions
\begin{align}
u(\tau,x_{\min}) &= \frac{\lambda_{B}\varepsilon_{h}}{r+\lambda_{\text{B}}+\lambda_{\text{C}}}\left(\e^{-(r+\lambda_{\text{B}}+\lambda_{\text{C}})\tau}-1\right),\label{e:x_min_BC}\\
\frac{\partial^{2} u}{\partial x^{2}}(\tau,x_{\max}) &= 0,\label{e:x_max_BC}
\end{align}
for all $\tau\in [0,T]$ completing the problem \eqref{e:PDE_u}.

\subsection{Numerical solution using finite differences}\label{sec:meth_FD}

The goal of this subsection is to introduce the approximate scheme for the transformed problem \eqref{e:PDE_u} completed by boundary conditions \eqref{e:x_min_BC},\eqref{e:x_max_BC} (see the commentary in Section~\ref{sec:meth_BC}). We intend to use the idea of the method of lines, i.e., a separate discretization in time $\tau$ and space $x$ (see, e.g., \cite[Chapter 13]{LeVeque07}). The whole problem is implemented in the software MATLAB, which possesses several useful solvers for the evolution in time. Hence, we will approximate the spatial part of the problem \eqref{e:PDE_u},\eqref{e:x_min_BC},\eqref{e:x_max_BC}, which results in the system of ODEs. Then we will use a suitable ODE solver in MATLAB to iterate it in time. The PDE of the problem \eqref{e:PDE_u} is typical advection--diffusion--reaction equation. Since the problem is situated in one spatial dimension, it is sufficient to use finite differences.

As we mentioned in the previous subsection, we work on the interval $[x_{\min},x_{\max}]$ in $x$. Let us discretize this interval by the uniform grid with constant step size $\delta_x = x_{i+1}-x_{i}$ for $i = 0,1,\ldots, N_x-1$, hence, there is $x_{0}=x_{\min}$ and $x_{N_x}=x_{\max}$. We denote the approximate value of $u(\tau,x_{i})$ by $U_{i}$ and the value of $f(V(\tau,x_{i}))$ by $F_{i}$. 

\begin{figure}[!ht]
\includegraphics[width=\textwidth]{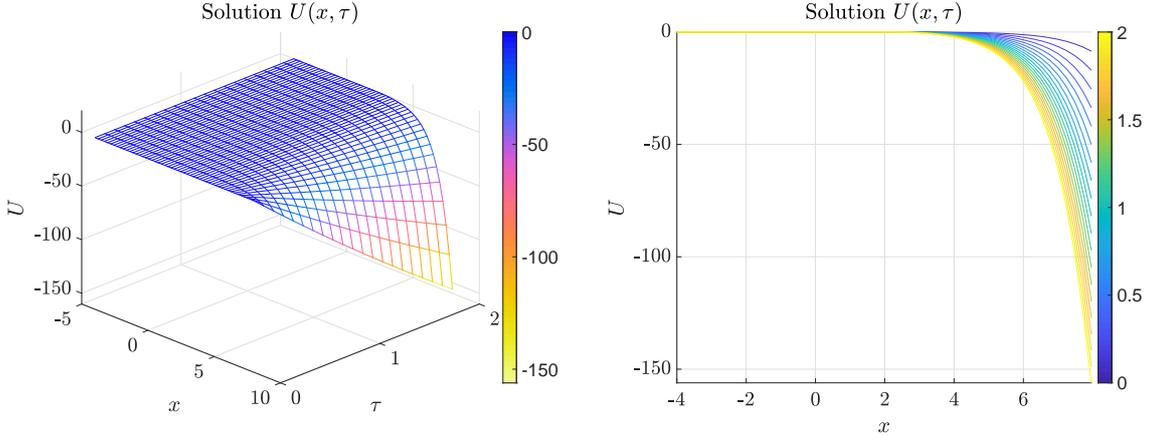}
\caption{Example of the solution computed by finite differences. The first picture contains the 3D plot of the solution in $(x,\tau)$. Contours of the solution for different values of $\tau$ are depicted on the right. The plotted solution was computed for the case $x_{\min} = -4$, $x_{\max} = 8$, and the strike $K = 15$ (the other parameters are stated in the next section).}\label{FIG:solution example}
\end{figure}

We acknowledge that it is quite common to use a non-uniform grid in $x$ as was presented in other works, e.g., \cite[Chap. 10]{Rouah13} or \cite[Sec. 4.2]{Hout17}, where the mapping function can be for example hyperbolic sine. One can then use very rough grid while maintaining reasonable density of points around the strike, where the solution changes most significantly. This approach can be useful if one aims to make large amount of numerical experiments (for different parameters of the model for example) with low computation cost. However, if we look at our problem from the point of view of numerical mathematics, the solution of \eqref{e:PDE_u} (see Figure \ref{FIG:solution example}) does not include any dramatic oscillations or boundary layers. Hence, we can achieve a reasonable solution with relatively rough uniform grid. Since our goal in this paper is to compare different approaches of solving the problem \eqref{e:PDE_u}, e.g., for different step sizes $\delta$, using the non-uniform grid is not necessary or even desirable. 
 
We approximate derivatives in \eqref{e:PDE_u} by
\begin{equation}
\frac{\partial^{2} u}{\partial x^{2}}(\tau,x_{i}) \approx \frac{U_{i-1}-2 U_{i} + U_{i+1}}{\delta_x^2} \quad \text{ and }\quad \frac{\partial u}{\partial x}(\tau,x_{i}) \approx \frac{U_{i+1} - U_{i-1}}{2\, \delta_x}
\end{equation}
for every $i=1,\ldots, N_x-1$. Both of these differences are central with the error $O(\delta_x^2)$. Let us mention here that since our problem is diffusion--dominated, the first order central difference approximating the advection term does not bring in any oscillations and thus no stabilization is necessary.

To deal with the boundary condition \eqref{e:x_max_BC} at $x_{\max}$, we employ the classical idea of a ghost point, i.e., we consider the point $x_{N_x+1} = x_{N_x} + \delta_x$ outside of the grid. {With this point, we can approximate the second derivative in \eqref{e:x_max_BC} again by the second order central difference, which yields
\begin{equation}\label{e:2nd_BC_approx}
\frac{U_{N_x-1}-2 U_{N_x} + U_{N_x+1}}{\delta_x^2} = 0.
\end{equation} 
Hence, we have the system of ODEs
\begin{equation}\label{e:ODE_system}
\frac{\partial U_{i}}{\partial \tau} = \frac{1}{2}\sigma^{2}\cdot\frac{U_{i-1}-2 U_{i} + U_{i+1}}{\delta_x^2} +\rho\cdot \frac{U_{i+1} - U_{i-1}}{2\, \delta_x}  -(r+\lambda_\text{B}+\lambda_\text{C})\cdot U_{i}-F_{i},\; \text{ for } i=1,\ldots,N_x-1.
\end{equation}
The equation \eqref{e:ODE_system} for $i=N_x$ contains the value $U_{N_x+1}$ outside of the grid, i.e., the value at the ghost point. Expressing $U_{N_x+1}$ from \eqref{e:2nd_BC_approx} and substituting it into \eqref{e:ODE_system} for $i=N_x$ give us the equation
\begin{equation}\label{e:x_max_ODE}
\frac{\partial U_{N_x}}{\partial \tau} = \rho\cdot \frac{U_{N_x} - U_{N_x-1}}{\delta_x}  -(r+\lambda_\text{B}+\lambda_\text{C})\cdot U_{N_x}-F_{N_x},
\end{equation}
for the computation of the value $U_{N_x}$ at the boundary $x_{\max}$.

We are using MATLAB's \texttt{ode15s} solver for the evolution of \eqref{e:ODE_system}, \eqref{e:x_max_ODE} in time. It is known that if the coefficient in front of the second derivative in \eqref{e:PDE_u} is small, the problem could be stiff. Hence, it is safer option to use \texttt{ode15s}, which is built to deal with stiff problems, rather than \texttt{ode45}. There are of course other options for the evolution in time, for example methods of alternating-direction implicit (ADI) type like Douglas method - see, e.g., \cite[Chap. 10]{Rouah13} or \cite[Chap. 13]{Hout17}. This type of methods is often used for problems in multiple spatial dimensions. Despite the fact, that our problem is in one spatial dimension, we could use ADI method to split the iterative process into three stages, the advection part, the diffusion part and the source part. This approach is very viable and would be better than for example Euler's method, lowering the restriction of the Courant–Friedrichs–Lewy (CFL) condition. However, \texttt{ode15s} solver is adaptive in time and very capable of dealing with this kind of problems, as well.

\subsection{Numerical solution using the transformation to heat equation}
\label{sec:meth_HE}

In order to evaluate the double integral in \eqref{e:sol_heat} numerically, a suitable quadrature has to be chosen. Although there exist many sophisticated quadratures, see for example monographs by \citet{Krylov62,Stoer02,Davis07,Dahlquist08}, for the clarity of our numerical comparison we use the simple trapezoid method. Let us consider an equidistant partition of the interval $[0,\tau]$ 
\[ 
s_i = i \cdot \delta_s, \quad i= 0, 1, \ldots, N_s 
\]
for given $\delta_s \in \mathbb{R}$ and $N_s \in \mathbb{N}$ such that $N_s \cdot \delta_s = \tau$. 
The inner integral needs to be restricted to a bounded interval in order for the trapezoid method to be applicable. 
Let $y_{\max}>0$ be sufficiently large number which shall be specified in the forthcoming sections and let $\delta_y \in \mathbb{R}$ and $N_y \in \mathbb{N}$ be such that $N_y \cdot \delta_y = 2 y_{\max}$. We then define the partition of the interval $[-y_{\max},y_{\max}]$ as 
\[
y_i = -y_{\max} + i \cdot \delta_y, \quad i = 0,1,\ldots, N_y.
\]
We can now express the trapezoid value as 
\begin{equation} \label{e:HE_trap_integral}
U_{\text{HE}}(\tau,x) = -\frac{1}{4\sqrt{\pi}} \sum_{i=0}^{N_s-1} \sum_{j=0}^{N_y-1} (g(s_{i+1},y_{i+1})-g(s_i,y_i)) \, \delta_y \, \delta_s, 
\end{equation}
in which
\[
g(s,y)= g(s,y;\tau,x) = \e^{-y^2-(r+\lambda_B+\lambda_C)(\tau-s)} f(v(s,x+\rho \tau + \sqrt{2(\tau-s)}\sigma y)).
\]

\subsection{Numerical solution using the Monte-Carlo method}
\label{sec:meth_MC}

Currently, the most widely used method to evaluate the formula~\eqref{e:sol_MC} with the subsequent expressions is the Monte-Carlo simulation. The incorporation of a change of variables and the simplification of the term $D_{\hat{r}}(t,u)$ in~\eqref{e:sol_MC} results in
\begin{align}
U_*(t,S) &= - \int_t^T \e^{-(r+\lambda_B+\lambda_C)(u-t)}\E[f_*(V(u,S(u))) \, | \, S(t) = S] \, \d u, \notag \\
&= - \int_0^{T-t} \e^{-(r+\lambda_B+\lambda_C)s} \E[f_*(V(t+s,S(t+s))) \, | \, S(t) = S] \, \d s, 
\end{align}
in which $* \in \{$ CVA, DVA, FCA, COLVA $\}$ and $S(t)$ is modelled as the geometric Brownian motion. After the change of variables \eqref{e:chov} we get
\begin{equation}
U_*(\tau,x) = - \int_0^\tau \e^{-(r+\lambda_B+\lambda_C)s}\E[f_*(v(T-\tau+s,\ln(S(T-\tau+s)))) \, | \, \ln(S(T-\tau)) = x] \, \d s. \label{e:MC_subs_individual_terms}
\end{equation}

We shall compute the integral term in~\eqref{e:MC_subs_individual_terms} again via the trapezoid method on the interval $[0,\tau]$ with partition
\begin{equation} \label{e:MC_time_partition}
s_i = i \cdot \delta_s, \quad i= 0, 1, \ldots, N_s 
\end{equation}
with $\delta_s \in \mathbb{R}$, $N_s \in \mathbb{N}$ and $\delta_s \, \cdot \, N_s = \tau$. 
The integral~\eqref{e:MC_subs_individual_terms} can be now computed as
\begin{equation}
\hat{U}_*(\tau,x) = - \frac{1}{2} \sum_{i=0}^{N_s-1} (g_*(s_{i+1})-g_*(s_{i})) \delta_s
\end{equation}
in which
\[
g_*(s_i) := g_*(s_i;x) = \e^{-(r+\lambda_B+\lambda_C)s} \varepsilon_*(x),  
\]
where $\varepsilon_*(x)$ is the Monte-Carlo estimate of the corresponding conditional expectation.
Finally 
\begin{equation} \label{e:MC_trap_integral}
U_{\text{MC}}(\tau,x) = \hat{U}_{\text{CVA}}(\tau,x) + \hat{U}_{\text{DVA}}(\tau,x) + \hat{U}_{\text{FCA}}(\tau,x) + \hat{U}_{\text{COLVA}}(\tau,x)
\end{equation}
yields the numerically computed value of~\eqref{e:sol_MC}. In practice, $\varepsilon_*(x)$ is usually efficiently estimated using the Quasi-Monte-Carlo method with low discrepancy Sobol sequences - see, e.g., \cite{Renzitti20}. The classical Monte-Carlo estimates will be calculated in the forthcoming section which is sufficient for the accuracy comparison purposes.

\section{Numerical results}\label{sec:results}

In this section we compare the pricing formula \eqref{e:sol_heat} with the Monte-Carlo pricing approach using the formula \eqref{e:sol_MC} and a numerical solution of the PDE \eqref{e:PDE_u}, namely with the solution obtained using the finite differences method. Also, we distinguish two collateral strategies: the un-collateralized contract, $X=0$, which leads to more intricate right source term of \eqref{e:PDE_U} compared to other choices and the delayed collateral, $X=V(t-t_0,S(t-t_0))$. We do not consider one-way CSA since the corresponding adjustments are quite similar to contracts with no collateral. We recall that the different collaterals were discussed in Section~\ref{sec:preliminaries}.

\subsection{Un-collateralized contract, \texorpdfstring{$X=0$}{X=0}}

We fixed parameters $\sigma = 0.25$, $q_s = 0.03$, $\gamma_s = 0$, $r=0.03$, $\lambda_B = 0.02$, $\lambda_C = 0.05$, $R_B = 0.4$, $R_C = 0.4$, $s_X = 0.012$, $X=0$, $\varepsilon_h$ as in~\eqref{e:varepsilon} and we valued the European call option with strike price $K=15$. All comparisons were made at $\tau = 2$, $S = \log{12}$.

For all the tests, we considered the pricing formula~\eqref{e:sol_heat} to be the benchmark solution which we computed with parameters $\delta_y = 2^{-3}$, $\delta_s = 2^{-10}$, $y_\mathrm{max} = 100$ (see Section~\ref{sec:meth_HE} for detailed explanation of these numerical parameters). These parameters seem to be the threshold ones such that finer discretization and larger integration region would not improve the accuracy significantly. 

The experiments pursued two main directions. First, we studied the behaviour of the finite difference scheme from Section~\ref{sec:meth_FD} while varying the spatial discretization $\delta_x$, the temporal discretization $\delta_\tau$ and the integration region $[x_\mathrm{min}, x_\mathrm{max}]$. The second experiment is focused on the Monte-Carlo formula~\eqref{e:MC_trap_integral} and its accuracy with respect to the number of simulated paths, denoted by $N$. 

While evaluating the performance of the finite difference scheme, we fixed $-x_\mathrm{min} = x_\mathrm{max} = 5$ and let $\delta_x = 2^{-2i}$, $i=1, \ldots, 4$, and $\delta_\tau = 2^{-2j}$, $j=1, \ldots, 5$, vary. The results are depicted in Figure~\ref{FIG:dx_dtau_test} which show a good agreement of the two approaches. As can be seen, the error is almost independent of the temporal discretization fineness which can be caused by the fact, that the \texttt{ode15s} solver chooses the time step adaptively.
\begin{figure}[!ht]
\centering
\begin{subfigure}[b]{.48\textwidth}
\centering
\includegraphics[width=\textwidth]{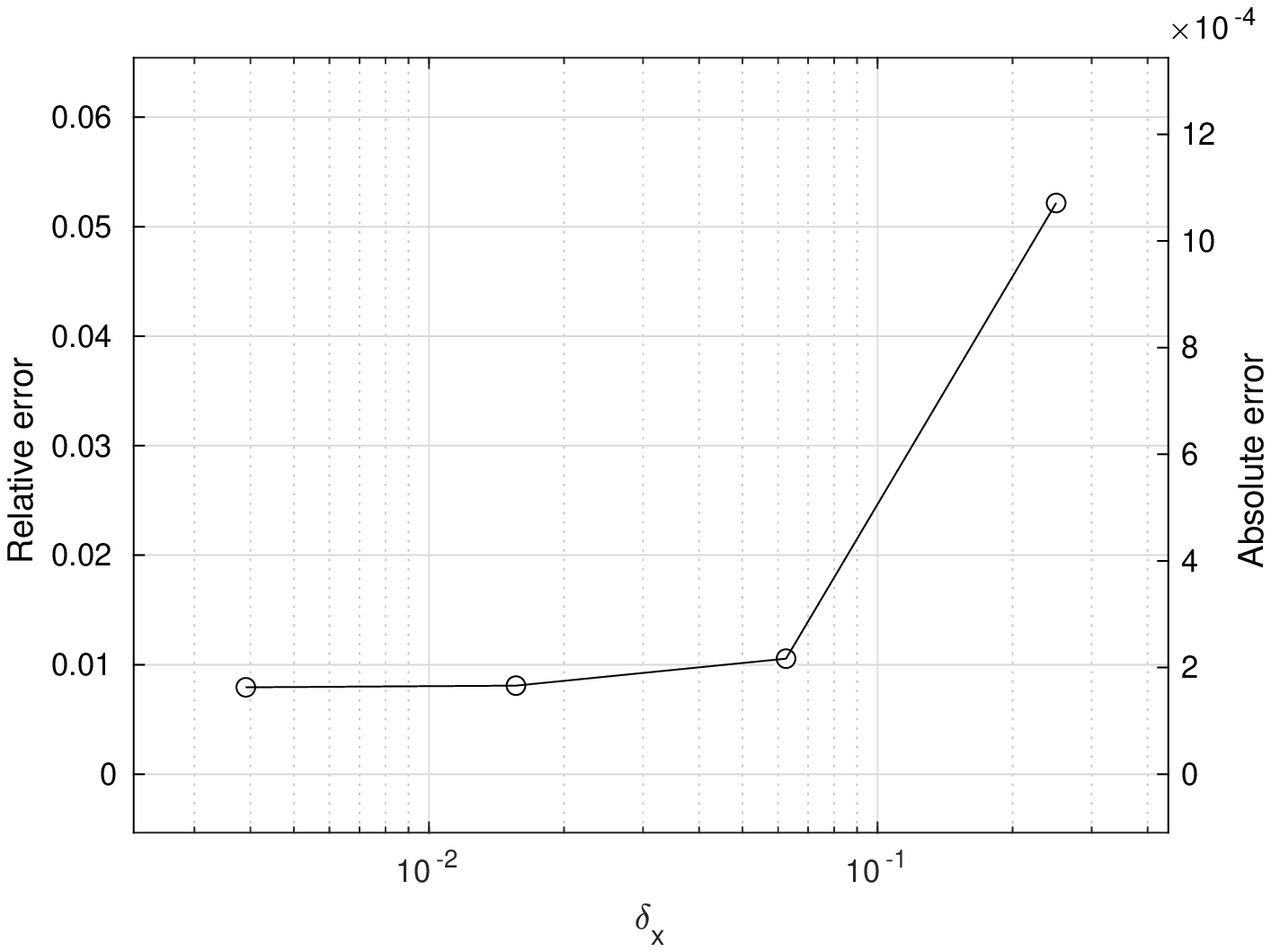}
\end{subfigure}
\begin{subfigure}[b]{.48\textwidth}
\centering
\includegraphics[width=\textwidth]{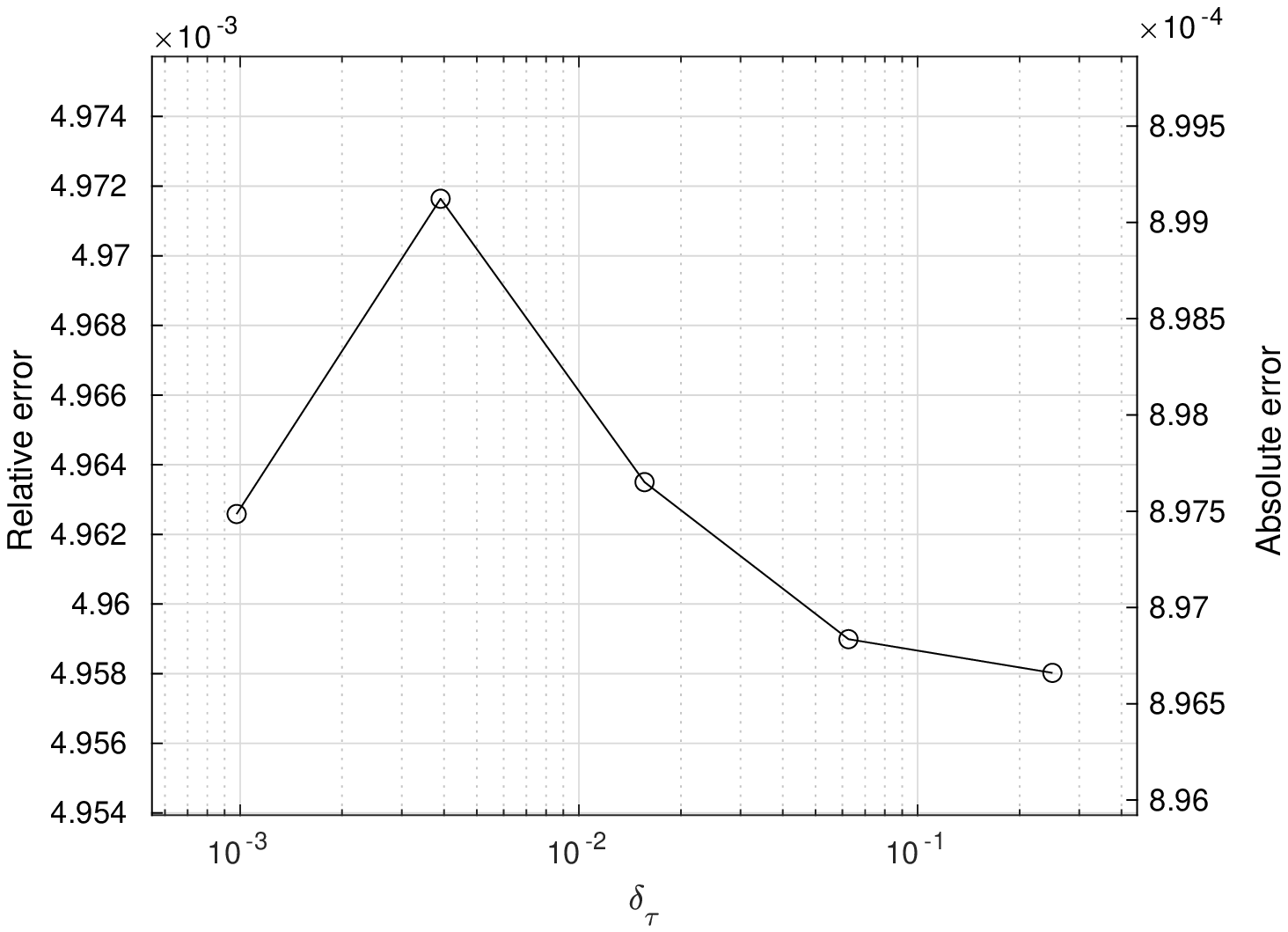}
\end{subfigure}
\caption{Comparison of finite difference solution (Section~\ref{sec:meth_FD}) and the heat equation formula (Section~\ref{sec:meth_HE}) at $\tau = 2$ and $S=\log 12$. The left panel captures the situation with fixed $\delta_\tau = 2^{-6}$ and the right panel depicts the case with fixed $\delta_x = 2^{-6}$. The weak dependence of the accuracy on the time step $\delta_\tau$ can be explained by the adaptive choice of the time step by the \texttt{ode15s} solver.} \label{FIG:dx_dtau_test}
\end{figure}

Next, we fixed $\delta_x = 2^{-6}$, $\delta_\tau = 2^{-3}$, $x_\mathrm{min} = -4$ and let $x_\mathrm{max} = 5, \ldots, 8$, vary. We do not vary $x_\mathrm{min}$, because $\e^{x_\mathrm{min}} = \e^{-4} \approx 0.018$ is sufficiently close to zero and taking more negative $x_\mathrm{min}$ would not provide much difference. As can be seen in Figure~\ref{FIG:xmax_test} the influence of the region truncation is notable for extremely small regions exclusively. Further enlargement does not significantly improve the accuracy. 

\begin{figure}[!ht]
\centering
\includegraphics[width=.7\textwidth]{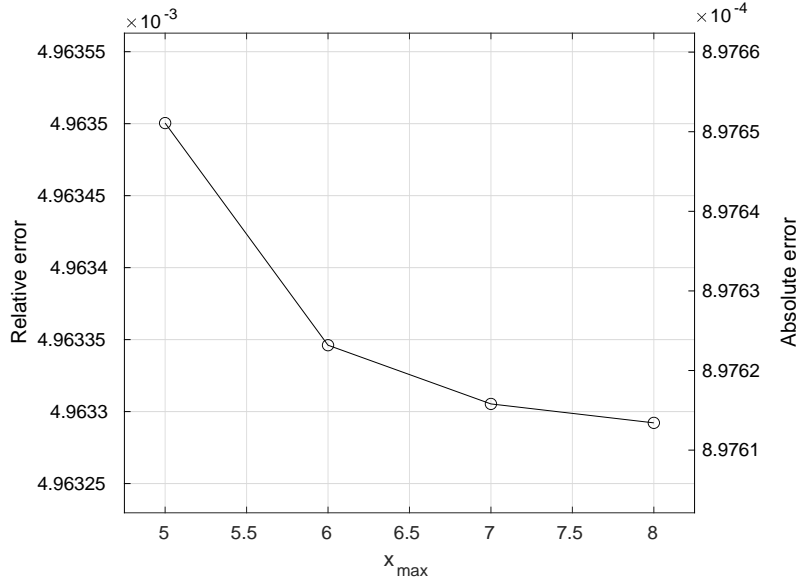}
\caption{Comparison of finite difference solution (Section~\ref{sec:meth_FD}) and the heat equation formula (Section~\ref{sec:meth_HE}) with $\delta_x=2^{-6}$, $\delta_\tau=2^{-6}$ and $x_\mathrm{max}$ variable.} \label{FIG:xmax_test}
\end{figure}

Finally, we set $\delta_x = 2^{-1}$, $\delta_\tau = 2^{-2}$, $-x_\mathrm{min} = x_\mathrm{max} = 5$ (note that these parameters influence only the number of points we use to compute the error since only the schemes~\eqref{e:HE_trap_integral},~\eqref{e:MC_trap_integral} were considered) and the Monte-Carlo time step $\delta_s = 2^{-10}$. The dependence of the error of the Monte-Carlo method on the number of simulated paths $N=10^{3+0.5j}$, $j = 0, \ldots, 4$ is depicted in Figure~\ref{FIG:N_test}.

\begin{figure}[!ht]
\centering
\begin{subfigure}[b]{.48\textwidth}
\centering
\includegraphics[width=\textwidth]{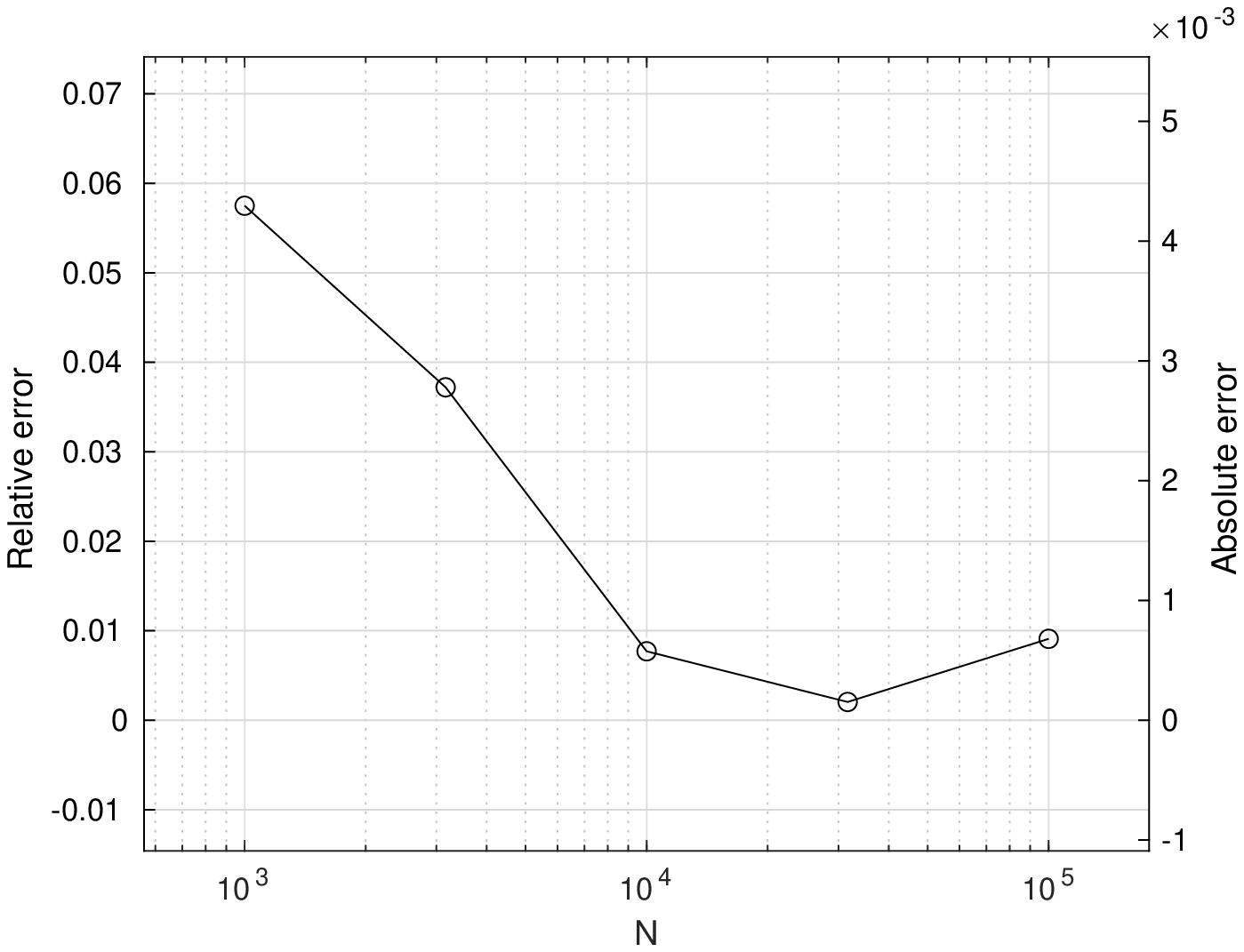}
\end{subfigure}
\begin{subfigure}[b]{.48\textwidth}
\centering
\includegraphics[width=\textwidth]{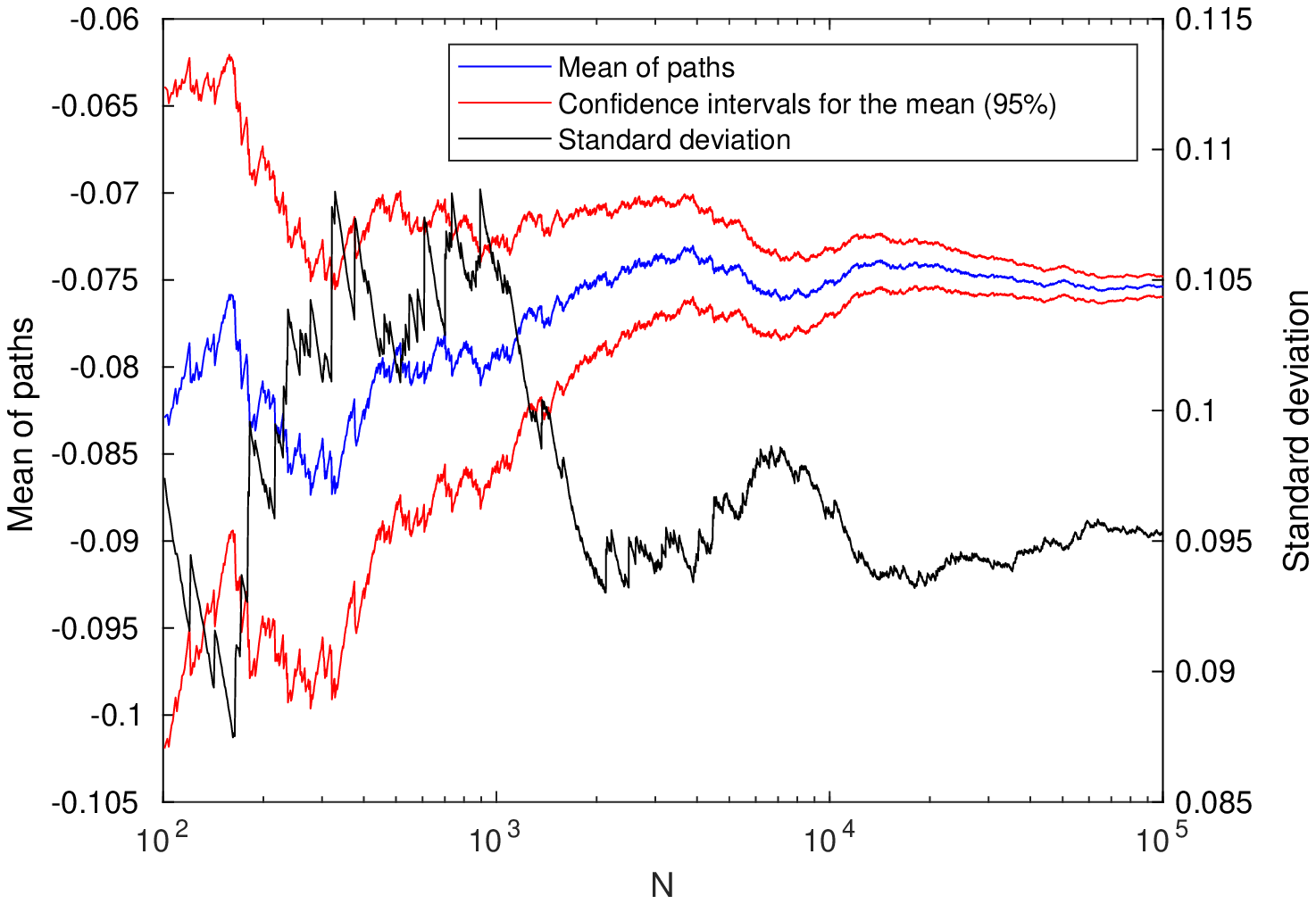}
\end{subfigure}
\caption{Comparison of the Monte-Carlo solution (Section~\ref{sec:meth_MC}) and the heat equation formula (Section~\ref{sec:meth_HE}) for variable number of simulated paths $N$.} \label{FIG:N_test}
\end{figure}

\subsection{Collateral as the delayed value, \texorpdfstring{$X=V(t-t_0,S(t-t_0))$}{X=V(t-t0,S(t-t0))}}

Let us now consider the collateral to be the delayed value as in \eqref{e:NPV} where $S(t-t_0)$ is approximated by its backward-discounted value, i.e.,
\[ S(t-t_0) \approx S(t) \exp\left\{-t_0(q_s-\gamma_s-\frac{1}{2} \sigma^2)\right\}. \] 
When dealing with one underlying only, Monte-Carlo simulation can be performed using the Brownian bridge pinned at $t$ and $t-t_0$ which leads to the same approximative underlying value at $t-t_0$. Also, one needs to adjust the time discretization in the computation of the integral appropriately for the partition points to hit exactly the past time $t-t_0$.

The model parameters are unchanged compared to the previous experiments except for $s_X = 0.02$. This is a necessary adjustment. Indeed, if we denote $X_1=V(t-t_1,S(t-t_1))$, $X_2 = V(t-t_2,S(t-t_2))$ for any delay terms $t_1, t_2 \in \mathbb{R}^+_0$ then since $V-X_i \leq 0$ for $i =1,2$ (for call options, BS price is increasing in $S$), the difference of the right hand side of~\eqref{e:PDE_U} can be expressed as
\[ f_1- f_2 = (- \lambda_B(1-R_B)+ s_X)(X_1-X_2). \]
If $\lambda_B(1-R_B) = s_X$ the values of the same derivatives are equal regardless of the delay $t_i$.

The delay term was set to be $t_0 = {10}/{252}$ (assuming $252$ trading days per year). The difference between the values of the derivatives with no delay ($X=V$) and with delay ($X=V(t-t_0,S(t-t_0))$) is depicted in the left panel of Figure~\ref{FIG:N_test_delayed}. The right panel shows the sample means and their sample standard deviations of the Monte Carlo simulations for varying number of simulations $N$. Confidence intervals (95\%) for the means are added assuming the number of simulations is large enough such that the means are normally distributed. 

\begin{figure}[!ht]
\centering
\includegraphics[width=.7\textwidth]{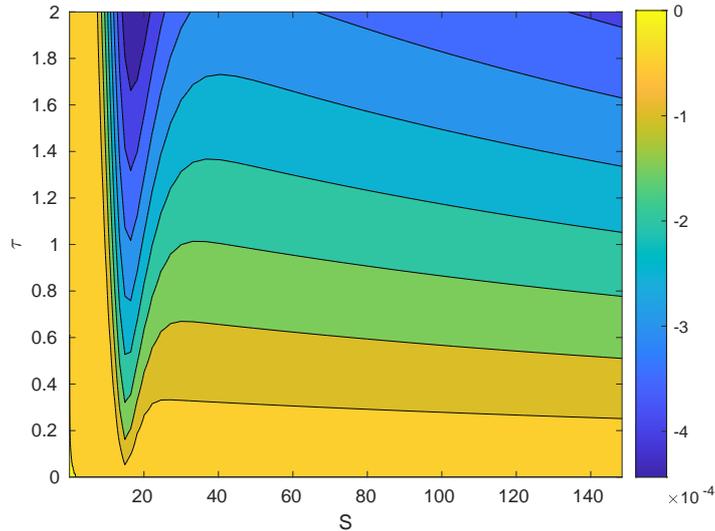}
\caption{Contour plot of the difference of the valued of the European call option with the undelayed collateral $X=V$ and the ten days delayed collateral $X = V(t-t_0,S(t-t_0))$ with $t_0 = {10}/{252}$.} \label{FIG:N_test_delayed}
\end{figure}

\section{Conclusion}\label{sec:conclusion}

The aim of this paper was to bring attention to analytical solutions of the partial differential equation used in value adjustment modelling. Although the PDE approach is known for more than a decade \citep{BurgardKjaer11PDE}, presented derivation of semi-closed formulas has not been published until now. When dealing with one underlying asset only, we can easily apply the transformation of the xVA PDE to the heat equation with known solution and hence obtain a fast and reliable semi-closed pricing formula that confirmed to be consistent with the Monte-Carlo simulations and with the numerical solution of the PDE obtained by the finite differences method. 

A special attention was paid to the example with collateral taken as the value from the past, typically ten days back is considered in practice. As a further research it might be interesting to study also what happens if the collateral is posted daily, weekly, monthly, etc. This however poses new analytical and numerical challenges, since the inhomogeneity $f$ is then discontinuous.

It is worth mentioning that xVA is often considered in a multi-dimensional setting, i.e., with multiple risk factors whose driving Wiener processes are instantaneously correlated. Such a setting then leads to a PDE with multiple spatial dimensions that could also be suited for transformation to heat equation as presented in this paper. A detailed study of the multi-dimensional setting was not in the scope of this paper but will be a subject of future research. Nevertheless, our analytic approach still might be of practical interest in the sense that one may need to study the valuation adjustment for each of the underlyings independently in order to identify their individual contribution to the risk. In this case, an analytic formula is definitely much more convenient than any Monte-Carlo simulation. On the other hand, calculating xVA for highly complex portfolio is nowadays performed only by means of Monte-Carlo and Quasi-Monte-Carlo simulations.

\section*{Funding}
  
The work was partially supported by the Czech Science Foundation (GA\v{C}R) grant no. GA18-16680S ``Rough models of fractional stochastic volatility''.

\section*{Acknowledgements}
  
Computational resources were supplied by the project "e-Infrastruktura CZ" (e-INFRA LM2018140) provided within the program Projects of Large Research, Development and Innovations Infrastructures.


{\small
\bibliographystyle{references/styles/jp+doi+mr+zbl}

}

\end{document}